\documentstyle[11pt,aaspp4]{article}
\eqsecnum
\catcode`\@=11
\def\gsim{\ifmmode{\mathrel{\mathpalette\@versim>}}
    \else{$\mathrel{\mathpalette\@versim>}$}\fi}
\def\lsim{\ifmmode{\mathrel{\mathpalette\@versim<}}
    \else{$\mathrel{\mathpalette\@versim<}$}\fi}
\def\@versim#1#2{\lower 2.9truept \vbox{\baselineskip 0pt \lineskip 
    0.5truept \ialign{$\m@th#1\hfil##\hfil$\crcr#2\crcr\sim\crcr}}}
\catcode`\@=12
\def\msun{\hbox{$M_\odot$}}
\def\teff{\hbox{$T_{\rm eff}$}}
\begin{document}

\title {Discovery of Extended Blue Horizontal Branches \nl
in Two Metal-Rich Globular Clusters
\altaffilmark{1}}

\author{R.\ Michael Rich\altaffilmark{2},
Craig Sosin\altaffilmark{5}, S.\ George Djorgovski\altaffilmark{3},
Giampaolo Piotto\altaffilmark{4},}
\author{Ivan R. King\altaffilmark{3},
Alvio Renzini\altaffilmark{6}, E.\ Sterl Phinney\altaffilmark{7},
Ben Dorman\altaffilmark{8,9},}
\author{James Liebert\altaffilmark{10}, and
Georges Meylan\altaffilmark{6}}

\altaffiltext{1}{Based on observations with the NASA/ESA
{\it Hubble Space Telescope}, obtained at the Space Telescope Science 
Institute, which is operated by AURA, Inc., under NASA Contract NAS
5-26555.}
\altaffiltext{2}{Columbia Astrophysics Laboratory, Columbia University, 
Mail Code 5242, New York, New York  10027;
rmr@astro.columbia.edu}
\altaffiltext{3}{Department of Astronomy, California Institute of Technology,
MS 105-24, Pasadena, CA 91125;
george@astro.caltech.edu}
\altaffiltext{4}{Dipartimento di Astronomia, Universit\`a di Padova,
Vicolo dell' Osservatorio 5, I-35122 Padova, Italy;
piotto@astrpd.pd.astro.it}
\altaffiltext{5}{University of California, Department of Astronomy,
601 Campbell Hall, Berkeley, CA 94720-3411;
csosin@aster.berkeley.edu, king@glob.berkeley.edu}
\altaffiltext{6}{European Southern Observatory,
Karl-Schwarzschild-Strasse 2, Garching bei M\"unchen, Germany;
arenzini@eso.org, gmeylan@eso.org}
\altaffiltext{7}{Theoretical Astrophysics, 
MS 130-33, California Institute of Technology,
Pasadena, CA 91125;
esp@tapir.caltech.edu}
\altaffiltext{8}{Laboratory of Astronomy and Solar Physics, Code 681,
NASA Goddard Space Flight Center, Greenbelt, MD 20771;
dorman@parfait.gsfc.nasa.gov}
\altaffiltext{9}{Astronomy Department, University of Virginia,
P.O.\ Box 3818, Charlottesville, VA 22903-0818}
\altaffiltext{10}{Steward Observatory, University of Arizona,
Tucson, AZ 85721; liebert@as.arizona.edu}

\newpage

\begin{abstract}
We have used WFPC2 to construct $B$, $V$ color--magnitude diagrams of
four metal-rich globular clusters, NGC 104 (47 Tuc), NGC 5927, NGC
6388, and NGC 6441.  All four clusters have well populated red
horizontal branches (RHB), as expected for their metallicity.
However, NGC 6388 and 6441 also exhibit a prominent blue HB (BHB)
extension, including stars reaching as faint in $V$ as the turnoff
luminosity.  This discovery demonstrates directly for the first time
that a major population of hot HB stars can exist in old, metal-rich
systems. This may have important implications for the interpretation
of the integrated spectra of elliptical galaxies.

The cause of the phenomenon remains uncertain.  We examine the
possibility that NGC 6388 and 6441 are older than the other clusters,
but a simple difference in age may not be sufficient to produce the
observed distributions along the HB.  The high central densities in
NGC 6388 and 6441 suggest that the existence of the blue HB (BHB)
tails might be caused by stellar interactions in the dense cores of
these clusters, which we calculate to have two of the highest
collision rates among globular clusters in the Galaxy.  Tidal
collisions might act in various ways to enhance loss of envelope mass,
and therefore populate the blue side of the HB.  However, the relative
frequency of tidal collisions does not seem large enough (compared to
that of the clusters with pure RHBs) to account for such a drastic
difference in HB morphology. While a combination of an age difference
and dynamical interactions may help, {\it prima facie} the lack of a
radial gradient in the BHB/RHB star ratio seems to argue against
dynamical effects playing a role.
\end{abstract}


\newpage

\section{ Introduction}

It is well known that a primary strong correlation exists between the
color distribution of the horizontal-branch (HB) stars in Galactic
globular clusters and their metallicities:\ metal-poor clusters have
essentially blue HBs (BHB) while metal-rich clusters have stubby red HBs (RHB).
Hot HB stars arise as a result of mass loss in RGB stars, and any
process that reduces the mean envelope mass either before or during the HB 
phase can also give rise to a BHB.  For example, high metallicity may promote
a hot extension of the HB in very old stellar populations if mass loss
increases with metallicity (cf.\ Greggio \& Renzini 1990).  A higher age
for the cluster will reduce both the main-sequence turnoff and the HB
mass, thus making the production of BHB stars more likely.
Also, dynamical interactions 
in dense clusters
might favor the production of BHB stars (cf.\ Fusi Pecci et al.\
1993).  The first option may be related to the increase in the UV flux
towards shorter wavelengths for $\lambda\lsim 2000$ \AA\ in some
elliptical galaxies (Code \& Welch, 1979; Bertola et al.\ 1980). The hot
stars responsible for the UV rising flux may indeed be HB stars that have
somehow managed to lose a major fraction of their envelopes, thus
spending their HB lifetime at high effective temperatures (Greggio \&
Renzini 1990).  While there is some circumstantial evidence to support
this idea (e.g.\ Ferguson \& Davidsen 1993; Dorman, O'Connell, \& Rood
1995), the actual presence of a significant proportion of hot HB stars
in an old and metal-rich population had not been observed to date.

In all such populations where individual stars are resolved, the HB
stars are indeed grouped in a red clump.  For example, HST observations
of the old, metal-rich clusters NGC 6528 and NGC 6553 have shown that
these bulge clusters have a purely RHB (Ortolani et al.\ 1995).
However, these two clusters might not be fully representative of all
possible varieties of metal-rich populations: Rich, Minniti, \& Liebert
(1993) obtained IUE spectra of a sample of metal-rich ([Fe/H] $>-0.6$)
clusters including NGC 6637, 6441, and 6388, finding weak but
significant UV fluxes. Also noteworthy is the discovery by Kaluzny \&
Udalski (1993) and Liebert, Saffer, \& Green (1994) of hot HB stars in
the old open cluster NGC~6791, which is considerably more metal-rich
at [Fe/H] $= +0.23, $ although it is younger and lower in stellar
density compared with the globular clusters.

Some globular clusters have a bimodal HB temperature distribution,
with a separate group of very hot HB stars coexisting with a cooler HB
component.  The first such cases to be noticed included NGC 1851 and NGC
2808.  These clusters have exceptionally high central stellar densities,
which suggested the possibility of a dynamical origin of the BHB
stars in these clusters, as extra mass loss during the red giant branch
(RGB) phase may be promoted by tidal encounters (Renzini 1983, 1984).  
Some of these clusters
(e.g.,\ M15) have color--magnitude diagrams (CMD) that are identical to those of
other low-density clusters of the same metallicity, except for the
addition of a hot HB component that is absent in the low-density
clusters (Buonanno et al.\ 1985).  Fusi Pecci et al.\ (1993) found a
correlation between cluster density and the presence of a hot HB
population and its extension, in particular for the
intermediate-metallicity clusters; but their sample did not include
clusters with [Fe/H] $>-0.6$.  Djorgovski et al.\ (1991) measured radial
color gradients in some high-concentration clusters, which in some (but
perhaps not all) cases could be traced to the gradients in the ratio of
BHB to RGB stars, thus strengthening the case for a possible
connection between the presence of hot HB stars and the dynamical state
of the clusters.  There is on the whole a growing body of evidence that
cluster dynamics, through close stellar encounters and/or formation of
binaries, can significantly alter the stellar populations in dense
stellar systems; for a review and further references, see Djorgovski \&
Piotto (1993).

In this {\sl Letter} we report the discovery of a BHB extension
in the metal-rich globular clusters NGC 6388 and NGC 6441, using the WFPC2
on board the {\sl Hubble Space Telescope} (HST).

\section{Observations and Color-Magnitude Diagrams}

The relevant properties of the program clusters are given in Table 1,
together with those of four additional metal-rich clusters also observed
with the refurbished HST but not included in the present program.  
Most of the
data in the first 6 columns are from the compilation by Djorgovski
(1993).  All the listed clusters have quite similar characteristics,
except that NGC 6388 and 6441 have high velocity dispersions (18 km
s$^{-1}$; Pryor \& Meylan 1993) and very high central surface brightnesses.  
The cluster
metallicities have been determined from the Ca II triplet method of
Armandroff \& Zinn (1988) to be $\rm [Fe/H]= -0.60 \pm 0.15$ for NGC
6388 and $-0.53 \pm 0.11$ for NGC 6441.

As part of a survey of globular cluster cores, NGC 6441 and NGC 6388
were observed by the WFPC2 on the HST on
12 September 1995, and on 25 February 1996, respectively.  Several
exposures were taken through the F555W and F439W filters, which
approximate Johnson $V$ and $B$.  These exposures were planned to reach
slightly fainter than the main-sequence turnoff; total integration 
times were 62 s
(NGC 6388) and 64 s (NGC 6441) in F555W, and 370 s in F439W for both
clusters.  Longer exposures were taken using the F218W filter.  We will
report on the analysis of the UV data at a later time.

The images were processed by the standard HST pipeline, and were then
trimmed, cleaned of cosmic rays, and combined using standard IRAF tasks.
We then used DAOPHOT (Stetson 1987) to identify stars, and to measure
magnitudes using standard PSF-fitting techniques.  More details of the
reduction procedure for all clusters observed in our HST program will be
presented in forthcoming papers.  The photometry was calibrated to the
WFPC2 instrumental system using the zero points given by Holtzman et al.\
(1995).  The F439W and F555W magnitudes were then converted to Johnson
$B$ and $V$, using transformations given in that same paper.

Figure 1 shows the color-magnitude diagrams of NGC 6388 and NGC 6441
compared with those of 47 Tuc and NGC 5927. While the two latter
clusters have only the classic RHBs typical of metal-rich globular clusters, 
NGC 6388 and 6441 have additional extended
BHBs.  The two BHBs have identical ridge lines when the CMDs are
shifted to identical distance modulus and reddening. 

The temperatures reached by the BHB stars are so high that their
$V$-band magnitudes are determined by the large bolometric corrections
(several magnitudes), and the BHB becomes a {\it vertical branch} in these
CMDs, reaching below the
main-sequence turnoff.  Note that a hint of a blueward extension of the HB
of NGC 6388 was present already in a ground-based CMD
of this cluster (Silbermann et al.\ 1994), but that photometry
was too shallow to reach even the bright top of the vertical branch.
Nevertheless, this suggests that the blue extension may be present at
even larger distances from the cluster center than those explored by our
WFPC2 frames.

Other metal-rich globular clusters known to have HST CMDs
include NGC 6528 and 6553 (Ortolani et al.\ 1995) and NGC 6624 (Sosin \&
King 1995).  All these clusters have RHBs similar to that of 47 Tuc and NGC
5927. Hence, NGC 6388 and NGC 6441 stand out as the only
metal-rich clusters with prominent BHB populations.  Since
their metallicities are comparable to those of the other clusters, with
normal RHBs, we are forced to conclude that this is yet another
manifestation of a second-parameter effect, by which we mean that
metallicity is not the only factor which governs the morphology of the HB.

\section{Discussion}

The presence of the classical second-parameter effect, i.e., a mostly red HB 
in a metal-poor cluster, is well established among clusters in the outer
halo.  However, Lee, Demarque, \& Zinn (1994) found no evidence for a
second-parameter effect among clusters with galactocentric distance less
than $\sim 8$ kpc.  Our findings clearly demonstrate that
second-parameter effects {\it are} actually present among Galactic-bulge 
globular
clusters, although the physical origin of the effect may be different in
the bulge and in the halo.  Given that the clusters in Table~1 with and
without BHB extensions do not differ substantially in their metallicities, 
we consider whether either age differences or stellar interactions can be
responsible for the extended HBs in NGC 6388 and 6441.

The effective temperature of Pop~II stars burning helium in their cores
is primarily a function of how much envelope has survived mass loss on
the RGB; see for example the extensive grids of HB models by Sweigart \&
Gross (1976) or Dorman, Rood, \& O'Connell (1993).  In principle, all
stellar populations develop a BHB when they become sufficiently old,
as stars with lower and lower mass evolve off the main sequence, and
therefore reach the HB with a lower and lower envelope mass.  However,
the age at which this happens is a function of metallicity, helium
enrichment, and mass loss during the RGB phase (e.g.,\ Greggio \& Renzini
1990).  In turn, mass loss likely depends on composition, but we have
neither an observational nor a theoretical relationship for this
dependence, which prevents {\it a priori} predictions of the HB
morphology.

At high metallicity, an increase in age can thus populate a blue tail
with stars at the low end of the mass distribution (see Renzini 1977,
Figure 3.1) while the dominant HB remains red. At the composition [Fe/H]
$=-0.47$, [O/Fe] $=+0.23,$ the nearest in the Dorman et al. 
(1993) model grid, the difference in envelope mass between ZAHB stars at
5000K and 11,500K (approximately the difference in $\teff$
between stars on the red and blue sides of the HB) is only $0.04
M_\odot,$ which would correspond to an age difference of $\sim 4$ Gyr,
the two clusters with hot HB extensions being older.
However, it is difficult to prove whether such an age difference alone can 
transform a stubby red HB such as that of 47 Tuc into a HB such as that 
exhibited by NGC 6388 or 6441. There is indeed a hint for a mildly bimodal 
distribution of stars on the HBs of these two clusters (cf. Figure 1), which 
may require a bimodal distribution of HB masses. 
Such bimodality is not
required to account for the HB of 47 Tuc, although it may be hidden by the 
very weak dependence of ZAHB temperature on envelope mass for red HB stars.
We conclude that further study is required to assess whether an older age 
can account for the peculiar HBs of NGC 6388 and 6441. If these
clusters are indeed older, then they would also have redder, fainter
turnoffs, but deeper HST observations are needed for this test.

We now turn to a second option, i.e., to the possible dynamical origin
of the hot HB component in these clusters. Again, the average extra mass loss
caused by environmental effects must exceed $\sim 0.04\,\msun$ in order
to generate blue HBs such as those shown in Figure\ 1.  Tidal stripping of
part of the envelope while stars are ascending the RGB is one possibility.
However, other more complicated alternatives can be envisaged
besides direct stripping at the time of the tidal collision.  For
example, orbital angular momentum can be transferred in a close flyby,
and stored in a red-giant envelope. In turn, this might promote enhanced
mass loss along the remaining fraction of the RGB phase, or even deep
mixing between the envelope and the upper part of the hydrogen-burning
shell (Sweigart 1997a,b).  Unfortunately, hydrodynamical phenomena such as
red-giant mass loss and mixing are exceedingly difficult to model, hence
these possibilities remain conjectural at this stage.

There are, however, established correlations between the blue extent of the HB
and cluster density or concentration.  Fusi Pecci et al.\ (1993) find the
strongest correlation of their blue tail length parameter (BT) with
[$c -2M_{\rm V}$], where $c$ is the King cluster concentration parameter and
$M_{\rm V}$ is the cluster integrated absolute magnitude.  However, while 
[$c-2 M_{\rm V}$] = 20.8 and 20.6 for NGC 6388 and 6441, respectively, 
[$c -2M_{\rm V}$] = 21.0 in the
pure-RHB cluster 47 Tuc. Still, we note that NGC 6388 and 6441
stand out as having some of the highest values of the central surface 
brightness and velocity dispersion 
among globular clusters in the Galaxy (cf. Pryor \& Meylan 1993;
Djorgovski 1993; see also Table 1), indicating very high central stellar densities, 
hence high rates of stellar interactions.  
It is of further interest to note that the globular cluster
G1 (=Mayall II) in M31 has a modest blueward HB extension to its dominant
RHB (Rich et al.\ 1996). This cluster has a central surface brightness as high 
as that of NGC 6388 and 6441, and a metallicity comparable to that of 47 Tuc.

It is reasonable to assume that in a
cluster the frequency of very close encounters
is some multiple of the frequency of stellar physical collisions,
for which early estimates were made by Hills \& Day (1976).
King (1997) shows that the rate of stellar collisions in a
King-model globular cluster is $\sim 5\times 10^{-15}
(\Sigma_0^3r_c)^{1/2}$, where $\Sigma_0$ is the central surface
brightness in the units of $L_{\odot\rm V}\>{\rm pc}^{-2}$ (equivalent to
$\mu_{\rm V}=26.41$), and $r_c$ is the core radius in pc.  (This formula was
derived by integrating the local collision rate over a density
distribution that is a good approximation for King models.  It is still
approximate, however, in that all stellar masses were taken to be
$0.4\,\msun$ and all stellar radii $0.5\,R_\odot$.)

Column 9 (the last column) gives the total number of the stars appearing
in our WFPC2 fields that should have undergone a collision,
relative to 47 Tuc.  We obtained these values by dividing the
numbers in Column 8 by the luminosity of the core of each cluster,
then multiplying by the number of giants in our WFPC2 image of that
cluster.
It appears that NGC 6388 and 6441 again stand out with the
highest relative collision rates, but the difference with respect to the other 
clusters seems too small to account for the large difference in HB morphology.
In NGC 6388, for example, we see about 180 BHB stars, and one would therefore 
expect to see about 75 of them in 47 Tuc, but we see none in Figure 1. 
Therefore,
it appears difficult for dynamical processes alone to account for the observed 
effect. Quite possibly, besides dynamics another factor is at work,
for example age.
Perhaps tidally stripped stars exist also in 47 Tuc, but the cluster is not
old enough for them to have started to populate the BHB. 

Another problem for the close-encounter hypothesis comes from
considering the radial distributions of the BHB stars.
Since most tidal encounters will
take place within the dense core of the cluster, one would expect 
the BHB stars to be more centrally concentrated than the RHB stars. However,
neither in NGC 6388 nor in NGC 6441 is such an effect present: we find that 
BHB and RHB stars have just the 
same radial distribution within the statistical uncertainty.
The only group of stars with a markedly different radial
distribution are the (centrally concentrated) blue stragglers. 

It might be argued that dynamical relaxation will quickly cause the encounter
products to diffuse out of the cluster center, thus smoothing out their
distribution.  To test this hypothesis we performed
Fokker--Planck simulations of the evolution of stellar orbits in model
clusters with the parameters of NGC 6388 and NGC 6441.  These show that
$10^8$~yr and $10^9$~yr after visiting the cluster core, stripped stars will 
still be concentrated within $2r_c$ and $4r_c$, respectively.
We find instead the BHB/RHB ratio to be flat out to at least $\sim 6r_c$.
Stripping stars at an earlier stage, while still on the main sequence,
would not help, as this will only lengthen their lifetimes rather than produce
core-helium-burning stars with less massive envelopes.  

A number of scenarios have been suggested to make BHB stars from binaries
(Bailyn 1995),
but these would also predict a central concentration of the BHB stars that
is not present in our data. Nonetheless, a binary-related origin remains
an intriguing possibility. Green, Saffer, \& Liebert (1997) find
evidence that some of the BHB stars in the old open cluster NGC 6791 are 
radial-velocity
variables, i.e., presumably binaries. Similarly, the field subdwarf B
stars show a very high frequency of binaries (Allard et al.\ 1994).

In conclusion,
we still lack a definitive explanation for the peculiar HBs in NGC 6388
and 6441.  There is a hint that dynamical effects, perhaps along with
an age difference, are at work. If stellar encounters play a role in
causing these peculiar HBs, then
other clusters of comparable surface brightness should also exhibit
the same peculiarity to some extent.
These may include HP 1, Liller 1, and NGC
6440, which all have very high surface brightness, 
and future HST imaging of them can test this hypothesis.
Whatever their cause, the existence of hot HB extensions
in metal-rich stellar populations will have broad implications
for our understanding of the evolution of globular cluster
stars, and could affect our interpretation of the spectra of
other old stellar systems, such as elliptical galaxies.

\acknowledgments

This work was supported in part by NASA grant GO-6095 from STScI (RMR,
SGD, IRK, CS), NASA grant NAG5-2756 (ESP), NASA grants NAG5-700 and
NAGW-4106 (BD), the Bressler Foundation (SGD), and Agenzia Spaziale
Italiana (GP, AR).

\begin{deluxetable}{rcccccccc}

\tablecaption{Relevant Physical Parameters of the Clusters in this
Study}
\tablehead{
\colhead{NGC} &  
\colhead{[Fe/H]} & 
\colhead{$M_V$} & 
\colhead{$(m-M)_0$} &
\colhead{ $\log r_c$ }& 
\colhead{$\mu_{0,V}$} & 
\colhead{$\sigma_{\rm obs}$} &
\colhead{$N_{\rm coll}$}   &
\colhead{$N_{\rm coll}$}   \\
\colhead{} & 
\colhead{} & 
\colhead{} & 
\colhead{} & 
\colhead{(pc)}  & 
\colhead{(mag/sq.\arcsec)} & 
\colhead{(km{\thinspace}s$^{-1}$)} &
\colhead{(total)} &
\colhead{(our field)}}
\startdata
 104 &  $-0.71$ &   $-9.42$ &   13.31 & $-0.56$ & 14.30 & $\phn9.8$  & $1.00$ &  $1
.00$ \\
5927 &  $-0.31$ &   $-7.94$ &   14.52 & $-0.01$ & 15.48 &   \nodata  & $0.37$ &  $0
.05$ \\
6388 &  $-0.60$ &   $-9.46$ &   15.21 & $-0.40$ & 13.45 &    $18.9$  & $3.89$ &  $2 
.41$ \\
6441 &  $-0.53$ &   $-9.30$ &   15.15 & $-0.46$ & 13.53 &    $17.6$  & $3.29$ &  $3 
.12$ \\
6528 &  $-0.23$ &   $-6.44$ &   14.10 & $-0.78$ & 14.70 &   \nodata  & $0.45$ & \nodata \\
6553 &  $-0.29$ &   $-7.56$ &   12.72 & $-0.25$ & 15.04 &   \nodata  & $0.52$ & \nodata \\
6624 &  $-0.37$ &   $-7.82$ &   14.54 & $-0.82$ & 14.45 & $\phn5.3$  & $0.61$ &  $0 
.85$ \\

\tablecomments{All data in this table are from Djorgovski (1993) and
Pryor \& Meylan (1993), except for the collision rates, and the core
radius of 47 Tuc (NGC 104).  For the latter we used the value given by
De Marchi \& Paresce (1996).  The collision rates in 
Column 8 are calculated using
the formula (King 1997) given in the text, and are normalized to that of
47 Tuc ($6.61\times 10^{-8} {\rm stars}\ {\rm yr}^{-1}$ for the entire
cluster).  NGC 6624 is the only cusped cluster in the sample.}

\enddata

\end{deluxetable}

\newpage

\begin{figure}
\plotone{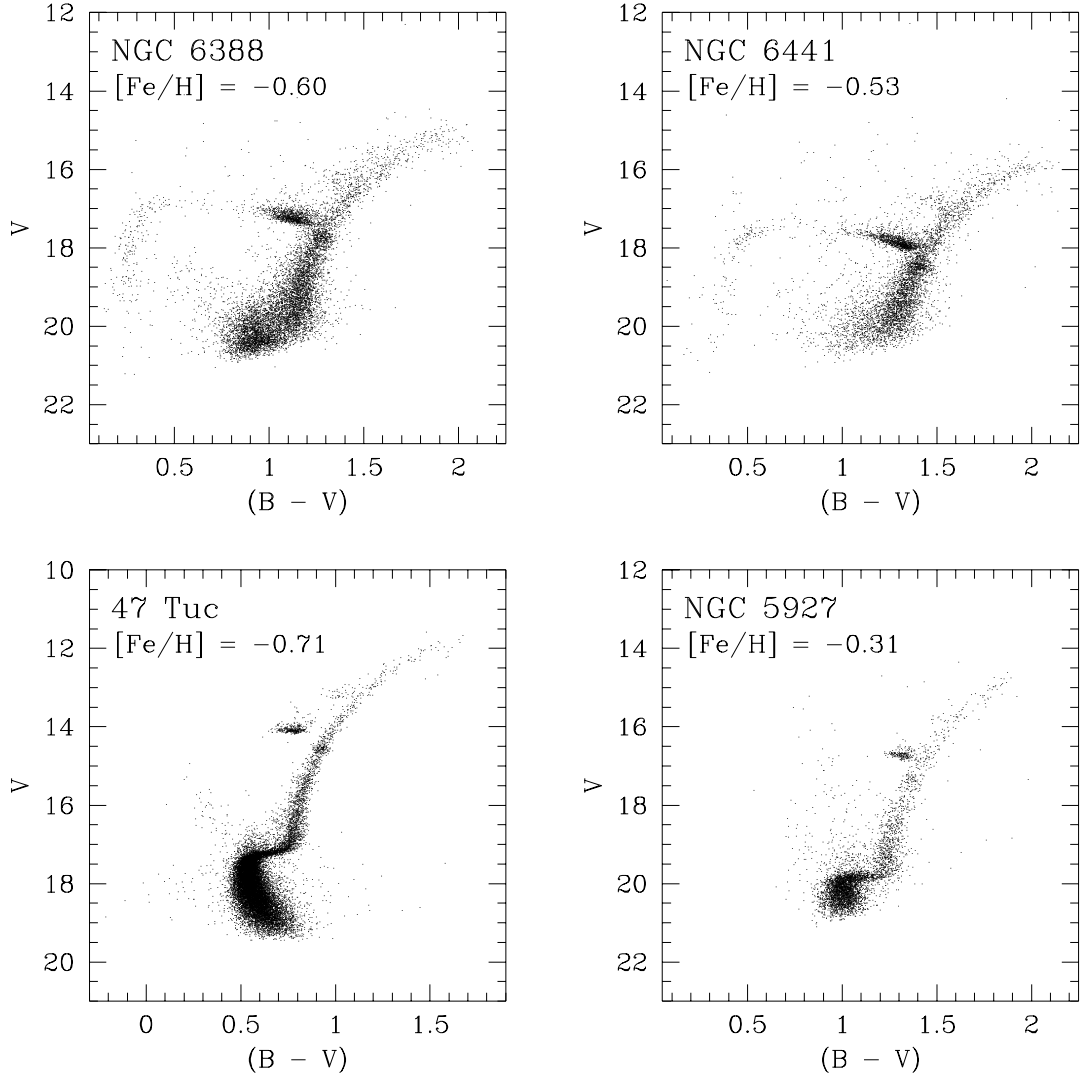}

\caption{Color-magnitude diagrams constructed from F439W ($B$) and
F555W ($V$) WFPC2 images of the globular clusters in this study.  The
red stubby horizontal branches of 47 Tuc and NGC 5927 are typical for
clusters of such high metallicity $(\rm [Fe/H]>-0.8)$. Extended blue
horizontal branches (such as those in NGC 6388 and 6441) have never
before been seen in such metal-rich globular clusters; they are the
only two cases now known.  Notice also that the morphology of the red
horizontal branches of NGC 6388 and 6441 is unusual compared to the
other clusters.}
\end{figure}


\newpage
\begin{references}

\noindent{Allard, F., Wesemael, F., Fontaine, G., Bergeron, P.,
\& Lamontagne, R.  1994, AJ, 107, 1565}


\noindent{Bailyn, C.D. 1995, ARAA, 33, 133}

\noindent{Bertola, F., Capaccioli, M., Holm, A.V., Oke, J.B. 1980, ApJ, 237,
L65} 

\noindent{Buonanno, R., Corsi, C.E., \& Fusi Pecci, F.  1985, A\& A, 145, 97}

\noindent{Code, A.D., \& Welch, G.A.  1979, ApJ, 228, 95}

\noindent{De Marchi, G., \& Paresce, F.\ 1996, ApJ, 468, L51}

\noindent{Djorgovski, S., Piotto, G., Phinney, E.S., \& Chernoff, D.F. 1991,
ApJ, 372, L41} 

\noindent{Djorgovski, S. 1993, in Structure and Dynamics of Globular Clusters,
ASPCS, 50, eds.\ S.\ G.\ Djorgovski \& G.\ Meylan (ASP: San Francisco),
p.\ 373}

\noindent{Djorgovski, S., \& Piotto, G. 1993, in Structure and Dynamics of
Globular Clusters, ASPCS, 50, eds.\ S.\ G.\ Djorgovski \& G.\ Meylan 
(ASP: San Francisco), p. 203} 

\noindent{Dorman, B., Rood, R.T., \& O'Connell, R.W. 1993, ApJ, 419, 592}

\noindent{Dorman, B., O'Connell, R.W., \& Rood, R.T.  1995, ApJ, 442, 105}

\noindent{Ferguson, H., \& Davidsen, A. 1993, ApJ, 408, 92}

\noindent{Fusi Pecci, F., Ferraro, F.R., Bellazzini, M., Djorgovski, S., Piotto,
G., \& Buonanno, R. 1993, AJ, 105, 1145} 

\noindent{Green, E. M., Saffer, R. A., \& Liebert, J. W. 1997, in 
``The Third Conference on Faint Blue Stars,'' eds. A. G. D. Philip,
J. W. Liebert, \& R. A. Saffer (Schenectady: L. Davis), in press}

\noindent{Greggio, L., \& Renzini, A. 1990, ApJ, 364, 35}


\noindent{Hills, J.G., \& Day, C.A. 1976, Ap.~Lett., 17, 87}

\noindent{Holtzman, J.A., et al. 1995, PASP, 107, 106}

\noindent{Kaluzny, J. \& Udalski, A. 1992, Act. Astr. 42, 29}

\noindent{King, I.R. 1997, Introduction to Stellar Dynamics, to be published
by University Science Books}

\noindent{Lee, Y.-W., Demarque, P., \& Zinn, R.J. 1994, ApJ, 423, 248}

\noindent{Liebert, J., Saffer, R.A., \& Green, E.M. 1994, AJ, 107, 1408}

\noindent{Ortolani, S., et al. 1995, Nat, 377, 701}

\noindent{Pryor, C., \& Meylan, G. 1993, in Structure and Dynamics of
Globular Clusters, ASPCS, 50, eds.\ S.\ G.\ Djorgovski \& G.\ Meylan
(ASP:\ San Francisco), p.\ 357}

\noindent{Renzini, A.\ 1977, in Advanced Stages in Stellar Evolution,
Saas--Fee, P.\ Bouvier \& A.\ Maeder (Geneva Obs: Sauverny), 
p.\ 151}

\noindent{Renzini, A. 1983, Mem. Soc. Astr. It. 54, 335}

\noindent{--------------. 1984, in Observational Tests of Stellar Evolution
Theory, ed. A. Maeder and A. Renzini (Dordrecht: Kluwer), p. 21}


\noindent{Rich, R.M., Minniti, D., \& Liebert, J. 1993, ApJ, 406, 489}

\noindent{Rich, R.M., Mighell, K.J., Freedman, W.L., \& Neill, J.D. 1996, AJ,
111, 768} 

\noindent{Silbermann, N.A., Smith, H.A., Bolte, M., \& Hazen, M.L. 1994, AJ,
107, 1764} 

\noindent{Sosin, C., \& King, I.R. 1995, AJ, 109, 639}

\noindent{Stetson, P.B. 1987, PASP, 99, 191}

\noindent{Sweigart, A.V. 1997a, ApJ, 474, L23}

\noindent{Sweigart, A.V. 1997b, in ``The Third Conference on Faint Blue 
Stars,'' eds.\ A.G.D.\ Philip, J.W.\ Liebert, \& R.A.\ Saffer
(Schenectady: L.\ Davis), in press}

\noindent{Sweigart, A.V., \& Gross, P.G. 1976, ApJS, 32, 367}

\end{references}
\end{document}